\documentclass{PoS}
 \usepackage{amssymb}
 \usepackage{amsmath}
 \usepackage{bbold}

\usepackage{slashed}

\newcommand{\ee}{m_\ell} 

\def\beq{\begin{equation}} 
\def\be{\begin{equation}}
\def\ee{\end{equation}}
\def\eq{\end{equation}}
\def\eeq{\end{equation}}
\def\beqa{\begin{eqnarray}}
\def\eqa{\end{eqnarray}}
\def\bea{\begin{eqnarray}}
\def\eea{\end{eqnarray}}
\def\beas{\begin{eqnarray*}}
\def\eeas{\end{eqnarray*}}
\def\beqas{\begin{eqnarray*}}
\def\eqas{\end{eqnarray*}}
\newcommand{\intu}{\int\limits_{-1}^{1}}
\newcommand{\intg}{\int\limits_{-1+\xi}^{1+\xi}}
\newcommand{\mpi}{m_\pi}
\newcommand{\Dp}{\Delta_\perp}
\newcommand{\MS}{M_S}
\newcommand{\HSn}{H_S^-}
\newcommand{\MIn}{M_1^-}
\newcommand{\MIIn}{M_2^-}
\newcommand{\MIIIn}{M_3^-}
\def\slashchar#1{\setbox0=\hbox{$#1$}
   \dimen0=\wd0
   \setbox1=\hbox{/} \dimen1=\wd1
   \ifdim\dimen0>\dimen1
      \rlap{\hbox to \dimen0{\hfil/\hfil}}
      #1
   \else
      \rlap{\hbox to \dimen1{\hfil$#1$\hfil}}
      /
   \fi}

\def\ps{\slashchar{p}}

\def\Ds{\slashchar{D}}
\def\ds{\slashchar{\partial}}

\def\Dpsab{\slashchar{\Delta}_{\perp \alpha \beta}}

\def\Ps{\slashchar{P}}

\def\As{\slashchar{A}}

\def\ns{\slashchar{n}}

\newcommand{\I}{\mathbb{1}}

\title{On higher twist chiral-odd pion generalized parton distributions }

\ShortTitle{Higher twist chiral-odd pion GPDs }

\author{B. Pire\\
CPHT, {\'E}cole Polytechnique, CNRS, 91128 Palaiseau, France}

\author{L. Szymanowski\\
        National Center for Nuclear Research (NCBJ), Warsaw, Poland}
        
\author{\speaker{S. Wallon}\\ Laboratoire de Physique Th\'eorique,  Universit{\'e} Paris-Sud, CNRS, 91405~Orsay, France {\em \&} \\
UPMC Univ. Paris 06, Facult\'e de Physique, 4 place Jussieu, 75252 Paris Cedex 05, France\\  \email{Samuel.Wallon@th.u-psud.fr}}

\abstract{We define in the framework of light-cone collinear factorization method, the chiral-odd generalized parton distributions (GPDs) of a pseudoscalar hadron (such as
the $\pi^0$) up to twist 6. For that, we 
introduce the relevant matrix elements for 2-parton non-local operators, as well as matrix elements for 3-parton non-local correlators. Their detailed parametrization is fixed based  on parity, charge conjugation and time reversal invariance. This leads to the introduction of 28 real GPDs, which are subject to constraints coming from the QCD equations of motion.
}

\FullConference{XXII International Workshop on Deep-Inelastic Scattering and Related Subject -DIS2014,\\
		April 28 - May 2, 2014,
		Warsaw, Poland}

\begin{document}

%
%
\maketitle
\thispagestyle{empty}
\renewcommand{\thesection}{\arabic{section}}

\renewcommand{\thesubsection}{\arabic{subsection}}
\noindent
\section{Introduction}
The higher twist extension  of the factorization properties of the leading twist amplitudes
for exclusive exclusive reactions in the generalized Bjorken regime~\cite{Ji:1998xh,Collins:1998be,Collins:1996fb} is a domain of intense recent research~\cite{Anikin:2000em, Anikin:2001ge, Anikin:2002wg, Anikin:2009hk, Anikin:2009bf, Anikin:2011sa, Braun:2011zr, Braun:2011dg}. We report here on an on-going study~\cite{Pire:2013vea} of pion GPDs in the framework of the light-cone collinear factorization (LCCF), 
the generalization of the Ellis--Furmanski--Petronzio (EFP) method~\cite{Ellis:1982cd, Efremov:1983eb, Teryaev:1995um} to the exclusive processes, which deals with the factorization in the momentum space around the dominant light-cone direction. Our first step  is to 
provide a classification of chiral-odd $\pi^0$ GPDs. We restrict ourselves to 2-parton and 3-parton correlators. We include the whole tower of twist
contributions from 2 to 6, but exclude any inclusion of pion mass effects, a question which has been successfully addressed recently~\cite{Braun:2012bg, Braun:2012hq}.

\section{Method}
\begin{figure}
\centerline{\includegraphics[width=14cm]{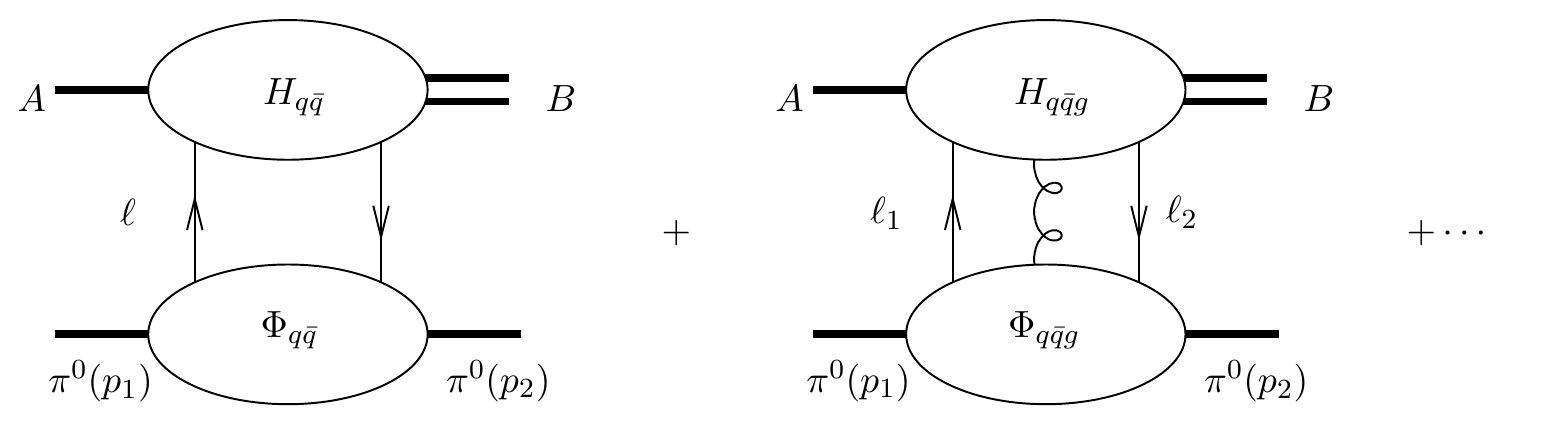}}
\caption{2- and 3-parton correlators attached to a hard scattering amplitude in the  process $A \,\pi^0 \to B \, \pi^0$.}
\label{Fig:NonFactorized}
\end{figure}

Let us consider a hard exclusive process. For definiteness, we name as $Q$ the involved hard scale (e.g. the $\gamma^*$'s virtuality in the case of deeply virtual Compton scattering (DVCS)).
We here recall the basics of the  LCCF in order to deal with  amplitude of exclusive
processes beyond the leading $Q$ power contribution. For definiteness, in view of
the next sections, we illustrate the key concepts for  
  the hard process $A \, \pi^0  \to B \, \pi^0$ (where $A$ and $B$ denote generic initial and final states in kinematics where a hard scale allows for a partonic interpretation, for example $A= \gamma^*$ and $B$ a particle with a scalar or a tensor coupling), written in
 the momentum representation and in the $n \cdot A =0$ light-like
gauge ($n^2=0$), as
\begin{eqnarray}
\label{GenAmp}
{\cal A}=
\int d^4\ell \, {\rm tr} \biggl[ H(\ell) \, \Phi (\ell) \biggr]+
\int d^4\ell_1\, d^4\ell_2\, {\rm tr}\biggl[
H_\mu(\ell_1, \ell_2) \, \Phi^{\mu} (\ell_1, \ell_2) \biggr] + \ldots \,,
\end{eqnarray}
where $H$ and $H_\mu$ are the coefficient functions
with two parton legs and three parton  legs,
respectively, as illustrated in figure~\ref{Fig:NonFactorized}.

The major step out of the leading twist approximation is to expand the hard part  $H(\ell)$  around
the dominant $p$ direction:
\begin{eqnarray}
\label{expand}
H(\ell) = H(y p) + \frac{\partial H(\ell)}{\partial \ell_\alpha} \biggl|_{\ell=y p}\biggr. \,
(\ell-y\,p)_\alpha + \ldots\,,
\end{eqnarray}
where $(\ell-y\,p)_\alpha = \ell^\perp_\alpha + (\ell \cdot p) \, n_\alpha$ allows one to extract higher twist contributions.
One can see that this procedure introduces
 a $\ell^\perp$ and a $\ell \cdot p$ dependence inside the hard part which does not seem to fit with 
the standard collinear framework. To proceed toward a factorized amplitude, one performs an
 integration by parts
to replace  $\ell^\perp_\alpha$ by $\partial^\perp_\alpha$ and $\ell \cdot p$ by $\partial \cdot p$
acting on the soft correlator in coordinate space.
 This leads to new operators
${\cal O}^\perp$ and ${\cal O}^-$ which contain
transverse derivatives, such as $\bar \psi \, \partial^\perp \psi $ and longitudinal derivatives\footnote{This completes the analysis performed in ref.~\cite{Pire:2013vea}.} along $n$ denoted as $\partial_n^{\gamma} \equiv (\partial \cdot p) n^\gamma$ such as $\bar \psi \, \partial_n^{\gamma}  \psi $
and thus
to the necessity of considering additional non-perturbative correlators
$\Phi^\perp $ and $\Phi^- $.

After these two steps, the amplitude takes the simple factorized form
\begin{eqnarray}
\label{GenAmpFac23}
\vspace{-.4cm}{\cal A}&=&
\int\limits_{-1}^{1} dy \,{\rm tr} \left[ H_{q \bar{q}}(y) \, \Gamma \right] \, \Phi_{q \bar{q}}^{\Gamma} (y)
+
\int\limits_{-1}^{1} dy \,{\rm tr} \left[ H^{(\perp,\,p)\mu}_{q \bar{q}}(y) \, \Gamma \right] \, \Phi^{(\perp,\,-)\Gamma}_{{q \bar{q}}\,\mu} (y) 
\nonumber \\
&+&\int\limits_{-1}^{1} dy_1\, dy_2 \,{\rm tr} \left[ H_{q \bar{q}g}^{(\perp,\,p)\mu}(y_1,y_2) \, \Gamma \right] \, \Phi^{(\perp,\,-)\Gamma}_{{q \bar{q}g}\,\mu} (y_1,y_2) + \cdots \,,
\end{eqnarray}
in which
the two first terms in the r.h.s correspond to the 
2-parton contribution and the last one to the 3-parton contribution.
As usual the antiquark contribution is interpreted as the
$[-1,0]$ part of this integral.

\section{Definitions}
Let us now construct the chiral-odd $\pi^0$ GPDs which parametrize the  2-parton and 3-parton correlators, taking into account constraints based on charge invariance, time invariance and parity invariance.
The 2-partons correlators may be written as
\beqa
\label{def-correlators-2-partons}
&&\langle \pi^0(p_2) | \bar{\psi}(z) \left[  \begin{array}{c}
                                  \sigma^{\alpha \beta} \\
				  \I \\
				  i \gamma^5  
                                 \end{array}
  \right] \psi(-z) | \pi^0(p_2) \rangle = \intu dx \, e^{i (x-\xi) P \cdot z + i (x+\xi) P \cdot z} \times\\
&& 
\left[  \begin{array}{ccc}
         -\frac{i}{\mpi} \left(P^\alpha \Dp^\beta -  P^\beta \Dp^\alpha  \right) H_T
\, & +  i \, \mpi \left(P^\alpha n^\beta -  P^\beta n^\alpha  \right) H_{T3}
&  - i \, \mpi \left(\Dp^\alpha n^\beta -  \Dp^\beta n^\alpha  \right) 
 H_{T4}
\nonumber \\
& \mpi \, H_S
\\
& 0
        \end{array}
  \right] \\
&&\hspace{2cm}  \mbox{twist  2 \& 4} \hspace{2.5cm} \mbox{twist  3} \hspace{3cm} \mbox{twist  4}
\nonumber
\eqa
where each (real) GPD depends on the arguments $x,\xi,t\,,$ and we underlined their twist content.
We note that due to ${\cal P}-$parity invariance, there is no twist 3 GPD associated with the $\gamma^5$ structure for the $\pi$ meson. This constraint will not survive in the nucleon GPD sector.

We now consider  correlators involving
the 3-parton and 2-parton (with transverse derivative). For the $\sigma^{\alpha \beta}$ structure, they read
\beqa
&&\hspace{-.3cm}\langle \pi^0(p_2) | 
       \bar{\psi}(z)  
\,\sigma^{\alpha \beta}\!
\left\{\!\begin{array}{c}
      i \stackrel{\longleftrightarrow}
{\partial_\perp^{\gamma}} \\
       g \, A^\gamma(y) 
                       \end{array}\!
 \right\} \!
 \psi(-z)| \pi^0(p_1) \rangle 
= \!\left\{\!
\begin{array}{l}\intu dx \, e^{i (x-\xi) P \cdot z + i (x+\xi) P \cdot z} \\
                          \int d^3 [x_{1,\,2,\,g}] \, e^{i P \cdot z (x_1+\xi) - iP \cdot y \, x_g + i P \cdot z \, (x_2 -\xi)}
                         \end{array}\!
\right\} \!\nonumber
\\
&& 
\label{def-correlators-3-partons-sigma-twist3}
\hspace{-.6cm}
\times \! \! \left[       i \, \mpi \left(P^\alpha g_\perp^{\beta \gamma} - P^\beta g_\perp^{\alpha \gamma}\right)\!
\left\{\begin{array}{c}
     T_1^T
\\
       T_1
       \end{array}
 \right\} +\frac{i}{\mpi} \left(P^\alpha \Dp^{\beta} - P^\beta \Dp^{\alpha}\right) \Dp^\gamma 
\left\{\begin{array}{c}
     T_2^T
\\
       T_2
       \end{array}\right\}  \ \ \ \ \ \ \mbox{(twist 3 \& 5)}\,\right.  \nonumber \\
&&
\label{def-correlators-3-partons-sigma-twist4}
\hspace{-.35cm}
\left. + i \, \mpi \left(\Dp^\alpha g_\perp^{\beta \gamma} - \Dp^\beta g_\perp^{\alpha \gamma}\right) 
\left\{\begin{array}{c}
     T_3^T
 \\
       T_3
       \end{array}
 \right\} 
 +  i \, \mpi \left(P^\alpha n^\beta -  P^\beta n^\alpha  \right) \Dp^\gamma \left\{\begin{array}{c}
     T_4^T
\\
       T_4
       \end{array}
 \right\} \right.\ \ \ \ \,  \mbox{(twist 4)} \nonumber \\
&&
\label{def-correlators-3-partons-sigma-twist5}
\hspace{-.35cm}
\left.  + i \, \mpi^3 \left(n^\alpha g_\perp^{\beta \gamma} - n^\beta g_\perp^{\alpha \gamma}\right)
\left\{\begin{array}{c}
     T_5^T
\\
       T_5
       \end{array}
 \right\} 
+i \mpi \left(n^\alpha \Dp^{\beta} - n^\beta \Dp^{\alpha}\right) \Dp^\gamma 
\left\{\begin{array}{c}
     T_6^T
\\
       T_6
       \end{array}\right\} 
\right]\,, \ \ \mbox{(twist 5)}
\eqa
where
\beq
\label{def-dx}
\int d^3 [x_{1,\,2,\,g}]
\equiv \intg d x_g \intu d x_1 \intu d x_2 \, \delta(x_g -x_2+x_1)\,,
\eq
and
$\stackrel{\longleftrightarrow}
{\partial_\perp^{\gamma}} \equiv \frac{1}2 (\stackrel{\longrightarrow}
{\partial_\perp^{\gamma}}-
\stackrel{\longleftarrow}
{\partial_\perp^{\gamma}})\,.$ 
The functions $T_i^T$ ($i=1, \cdots 6$) should be understood as $T_i^T(x,\xi,t)\,,$
while $T_i$ ($i=1, \cdots 6$) denotes $T_i(x_1,x_2,\xi,t)\,.$

For the $\I$ structure, the correlators are defined as
\beqa
\langle \pi^0(p_2) | 
       \bar{\psi}(z)
\,\I \!
\left\{\!\begin{array}{c}
      i \stackrel{\longleftrightarrow}
{\partial_\perp^{\gamma}} \\
       g \, A^\gamma(y) 
                       \end{array}\!
 \right\} \!
 \psi(-z)| \pi^0(p_1) \rangle  
&\!=& \!\left\{
\begin{array}{l}\!\intu dx \, e^{i (x-\xi) P \cdot z + i (x+\xi) P \cdot z} \\
                         \! \int d^3 [x_{1,\,2,\,g}] \, e^{i P \cdot z (x_1+\xi) - iP \cdot y \, x_g + i P \cdot z \, (x_2 -\xi)}
                         \end{array}
\right\} \nonumber
\\
\label{def-correlators-3-partons-id-twist4}
\hspace{2cm}
&\times&  
\mpi \, \Dp^\gamma
\left\{\begin{array}{c}
     H_S^{T4}
\\
       T_S
       \end{array}
 \right\}  \,. \hspace{1cm} \mbox{ (twist 4)}
\eqa

For the $i \gamma^5$ structure, the correlators read
\beqa
\langle \pi^0(p_2) | 
       \bar{\psi}(z)
\, i \gamma^5 \!
\left\{\!\begin{array}{c}
      i \stackrel{\longleftrightarrow}
{\partial_\perp^{\gamma}} \\
       g \, A^\gamma(y) 
                       \end{array}\!
 \right\} \!
 \psi(-z)| \pi^0(p_1) \rangle
&=& \left\{
\begin{array}{c}\intu dx \, e^{i (x-\xi) P \cdot z + i (x+\xi) P \cdot z} \\
                          \int d^3 [x_{1,\,2,\,g}] \, e^{i P \cdot z (x_1+\xi) - iP \cdot y \, x_g + i P \cdot z \, (x_2 -\xi)}
                         \end{array}
\right\} \nonumber
\\
\label{def-correlators-3-partons-igamma5-twist4}
\hspace{2cm}
&\times&  
 \mpi \, \epsilon^{\gamma \,  n \, P \, \Dp}
\left\{\begin{array}{c}
     H_P^{T}
\\
       T_P
       \end{array}
 \right\}  \,. \hspace{1cm} \mbox{ (twist 4)}
\eqa

We now consider  correlators involving
the 3-parton (with longitudinally polarized gluon) and 2-parton (with longitudinal derivative). We denote 
\beq
\label{def-dn-An}
\partial_n^{\gamma} \equiv (\partial \cdot p) n^\gamma \quad {\rm and} \quad A_n^{\gamma} \equiv (A \cdot p) n^\gamma\,.
\eq
For the $\sigma^{\alpha \beta}$ structure, they read
\beqa
&&\hspace{-.3cm}\langle \pi^0(p_2) | 
       \bar{\psi}(z)  
\,\sigma^{\alpha \beta}\!
\left\{\!\begin{array}{c}
      i \stackrel{\longleftrightarrow}
{\partial_n^{\gamma}} \\
       g \, A_n^\gamma(y) 
                       \end{array}\!
 \right\} \!
 \psi(-z)| \pi^0(p_1) \rangle 
= \!\left\{\!
\begin{array}{l}\intu dx \, e^{i (x-\xi) P \cdot z + i (x+\xi) P \cdot z} \\
                          \int d^3 [x_{1,\,2,\,g}] \, e^{i P \cdot z (x_1+\xi) - iP \cdot y \, x_g + i P \cdot z \, (x_2 -\xi)}
                         \end{array}\!
\right\} \!\nonumber
\\
&& 
\label{def-correlatorsN-3-partons-sigma-twist4et6}
\hspace{-.6cm}
\times \! \! \left[     
i m_\pi \left(P^\alpha \Dp^{\beta} - P^\beta \Dp^{\alpha}\right)n^\gamma 
\left\{\begin{array}{c}
     \MIn
\\
       M_1
       \end{array}\right\}  \ \ \mbox{(twist 4 \& 6)}\,\right.  \\
&&
\label{def-correlatorsN-3-partons-sigma-twist5}
\hspace{-.35cm}
\left. + \, i \, \mpi^3 \left(P^\alpha n^\beta -  P^\beta n^\alpha  \right) n^\gamma \left\{\begin{array}{c}
     \MIIn
\\
       M_2
       \end{array}
 \right\} \right.\ \  \mbox{(twist 5)}  
\left.  + \, i \, \mpi^3 \left(n^\alpha \Dp^{\beta} - n^\beta \Dp^{\alpha}\right) n^\gamma 
\left\{\begin{array}{c}
     \MIIIn
\\
       M_3
       \end{array}\right\} \
 \mbox{(twist 6)} \ \right]  \,. \nonumber
\eqa
For the $\I$ structure, the correlators are defined as
\beqa
\langle \pi^0(p_2) | 
       \bar{\psi}(z)
\,\I \!
\left\{\!\begin{array}{c}
      i \stackrel{\longleftrightarrow}
{\partial_n^{\gamma}} \\
       g \, A_n^\gamma(y) 
                       \end{array}\!
 \right\} \!
 \psi(-z)| \pi^0(p_1) \rangle  
&=& \!\left\{
\begin{array}{l}\!\intu dx \, e^{i (x-\xi) P \cdot z + i (x+\xi) P \cdot z} \\
                         \! \int d^3 [x_{1,\,2,\,g}] \, e^{i P \cdot z (x_1+\xi) - iP \cdot y \, x_g + i P \cdot z \, (x_2 -\xi)}
                         \end{array}
\right\} \nonumber
\\
\label{def-correlatorsN-3-partons-id-twist4}
\hspace{2cm}
&\times&  
\mpi^3 \, n^\gamma
\left\{\begin{array}{c}
     \HSn
\\
       \MS
       \end{array}
 \right\}  \hspace{1cm} \mbox{ (twist 5)}  \ .
\eqa

Altogether, the 2- and 3-parton correlators lead to the introduction of 28 different GPDs. They are not independent, and the reduction to an independent set is not a simple task. The two basic tools to implement this reduction is firstly the QCD equations of motion and secondly the constraints usually denoted as $n-$invariance \cite{Ellis:1982cd}.

\section{Constraints from QCD equations of motion}
 
We start with the Dirac equation for the quark field
\begin{eqnarray}
0 &=&\langle \pi^0(p_2) | \, (i \Ds \psi)_\alpha (-z) \, \bar{\psi}_\beta(z) \, |\pi^0(p_1) \rangle  \nonumber\\
&=&  \langle \pi^0(p_2) | \left[ i (\partial \cdot n) (\ps \psi)_\alpha (-z) \, 
 +i (\partial \cdot p) (\ns \psi)_\alpha (-z) \, 
 +(i \ds_\perp \psi)_\alpha (-z)\right. \nonumber \\
&& \left.+ \ g \, (\As_\perp \, \psi)_\alpha(-z) 
 + \ g \, (A \cdot p) (\ns \, \psi)_\alpha(-z)\right] \,\bar{\psi}_\beta(z) \, |\pi^0(p_1) \rangle  \,.  
 \end{eqnarray}
After a tedious but straightforward calculation of these five terms  and demanding the vanishing of the contributions multiplying
the four independent structures $\Ps_{\alpha \beta}\,,$ 
$\Dpsab\,,$
$\ns_{\alpha \beta}$ and\linebreak $i \, \epsilon^{\Dp P \,n \,\mu} \, \left(\gamma^5 \gamma_\mu\right)_{\alpha \beta}\,,$ we obtain the following four equations
\beqa
\label{I}
&&(x + \xi) (\mpi \, H_{T3} + \mpi H_S) + \frac{\Dp^2}{2 \mpi} H_T
+ 2 \, \mpi \, T^T_1 + \frac{\Dp^2}{\mpi}\, T^T_2  \nonumber\\
&&+\, \frac{1}2 \intg d x_g \intu dy \, \delta(x_g - x + y)\left(2 \, \mpi \, T_1(y,x) + \frac{\Dp^2}{\mpi} T_2(y,x) \right)=0 \,,
\eqa
\beqa
\label{II}
&&(x + \xi) \left(\frac{P^2}{\mpi} \, H_T - \mpi H_{T4} \right) + \mpi \left(-\frac{1}2 H_S + T_3^T + H_S^{T4}    \right) \nonumber \\
&&+\, \frac{\mpi}2 \intg d x_g \intu dy \, \delta(x_g - x + y) \left(T_3(y,x) + T_S(y,x) \right) \nonumber \\
&&
-\, \frac{\mpi}2 \intg d x_g \intu dy \, \delta(x_g - x + y) \, M_1(y,x) - \frac{\Delta \cdot p}{2 \mpi} H_T(x) - \mpi \, \MIn(x)
=0\,, \ \
\eqa
\beqa
\label{III}
&&(x + \xi) \, \mpi \, P^2 \, H_{T3}(x) + \frac{\mpi \, \Dp^2}2 \, H_{T4}(x) - 2 \mpi^3 \, T_5^T(x) - \mpi \, \Dp^2 \, T_6^T(x) \nonumber \\
&& 
- \frac{1}2 \intg d x_g \intu dy \, \delta(x_g + y - x) \left(2 \mpi^3 T_5(y,x) + \mpi \, \Dp^2 \, T_6(y,x)   \right) 
\nonumber \\
&&
-\, \frac{\mpi^3}2 \intg d x_g \intu dy \, \delta(x_g - x + y) \, \MS(y,x)
+ \, \frac{\mpi^3}2 \intg d x_g \intu dy \, \delta(x_g - x + y) \, M_2(y,x)
\nonumber \\
&&
- \frac{\Delta \cdot p}{2} \mpi H_{T3}(x) 
+ \frac{\Delta \cdot p}{2} \mpi H_{S}(x)
+ \mpi^3 \, \MIIn(x)
- \mpi^3 \, H_S^-(x)
=0\,,
\eqa
and
\beqa
\label{IV}
&&(x + \xi) \, H_{T4}(x) -\frac{1}2 H_{T3}(x) + T^T_4(x) + H_P^T(x) \nonumber \\
&&+ 
\frac{1}2\intg d x_g \intu dy \, \delta(x_g + y - x) \left(T_4(y,x) + T_P(y,x)     \right)\nonumber \\
&&
-
\frac{1}2\intg d x_g \intu dy \, \delta(x_g + y - x) M_1(y,x)      -\frac{\Delta \cdot p}{2 \mpi^2} H_T(x)
- \MIn(x)
=0\,.
\eqa
A second set of four equations is obtained in an analogous way by considering the various correlators involved in the following equation coming from the Dirac equation on the antiquark field 
\beq
\label{action-left}
0 = \langle \pi^0(p_2) | \, \psi_\alpha (-z) \, (i \Ds \bar{\psi})_\beta(z) \, |\pi^0(p_1) \rangle\,.
\eq
This new set of  equations  is related by charge conjugation to the previous ones.

\section{Constraints from the $n$-invariance}

The physical amplitude for the process 
$A \, \pi^0  \to B \, \pi^0$ should not depend on the arbitrary 
light-like vector $n$, which is involved when defining the Fourier transform with respect to the light-cone momentum 
direction $p$ in order to define the various GPDs, as well
as in the way one fixes the gauge. The requirement that $p \cdot n=1$ and $n^2=0$ effectively reduces to an invariance with respect to variations of $n_\perp$, i.e.
\begin{equation}
\label{eqAn}
\frac{d {\cal A}}{d n_\perp} =0 \,. 
\end{equation}

As shown in detail in refs.~\cite{Anikin:2009hk,Anikin:2009bf}, for the case of the $\gamma^* \to \rho$ impact factor, such a requirement can be reduced, at twist-3 level, to a set of equations relating the various distribution amplitudes (DA) involved.
Indeed,
after proper use of Ward identities, one can effectively factorize out the hard Born contribution from eq.~(\ref{eqAn}), in such a way that this equation leads to a set of constraints among the various non-perturbative correlators, i.e. {\it in fine} between the DAs themselves. Combining these equations with the one based on the QCD equations of motion, one can thus reduce the set of DAs to a minimal set, which are three independent DAs in the case of the twist 3 chiral-even $\rho$ meson. 

A similar analysis can be performed for any physical amplitude involving the chiral-odd $\pi^0$ GPDs. The detailed study of the corresponding reduction to a minimal set of independent chiral-odd GPDs for $\pi^0$ is under study.

\section*{Acknowledgements} 
This work is partly supported by the Polish Grant NCN
No. DEC-2011/01/B/ST2/03915, the French-Polish collaboration agreement 
Polonium,
the ANR ``PARTONS'', the PEPS-PTI ``PHENO-DIFF'', the Joint Research Activity Study of Strongly 
Interacting Matter (acronym HadronPhysics3, Grant Agreement n.283286) under the Seventh 
Framework
Programme of the European Community and by the COPIN-IN2P3 Agreement.


\end{document}